\documentclass[twoside,twocolumn]{IEEEtran}
\usepackage[T1]{fontenc}
\usepackage{color}
\usepackage{float}
\usepackage{amsmath}
\usepackage{amsthm}
\usepackage{amssymb}
\usepackage{graphicx}
\usepackage{ulem}
\usepackage{bm}
\usepackage{placeins}  
\usepackage{hyperref}  

\makeatletter

\floatstyle{ruled}
\newfloat{algorithm}{tbp}{loa}
\providecommand{\algorithmname}{Algorithm}
\floatname{algorithm}{\protect\algorithmname}

  \theoremstyle{plain}
  
  \theoremstyle{plain}
  
  \theoremstyle{plain}

\ifCLASSOPTIONcompsoc
\usepackage[caption=false,font=normalsize,labelfont=sf,textfont=sf]{subfig}
\else
\usepackage[caption=false,font=footnotesize]{subfig}
\fi

\usepackage{cite}
\usepackage{bm}
\usepackage{algorithmic}
\usepackage{algorithm}
\usepackage{graphicx}
\usepackage{bbm}
\usepackage{mathrsfs}
\IEEEoverridecommandlockouts
\makeatother

\providecommand{\propositionname}{Proposition}
\providecommand{\corollaryname}{Corollary}
\providecommand{\theoremname}{Theorem}

\begin{document}
\title{Batch Denoising for AIGC Service Provisioning in Wireless Edge Networks}
\author{Jinghang~Xu, Kun~Guo,~\IEEEmembership{Member,~IEEE}, Wei~Teng,~\IEEEmembership{Member,~IEEE}, Chenxi~Liu,~\IEEEmembership{Senior Member,~IEEE}, and~Wei~Feng,~\IEEEmembership{Senior Member,~IEEE}\thanks{J. Xu and K. Guo are with the School of Communications and Electronics Engineering, East China Normal University, Shanghai 200241, China (e-mail: 51275904080@stu.ecnu.edu.cn, kguo@cee.ecnu.edu.cn)  \textit{(Corresponding author: Kun Guo.)}}
\thanks{W. Teng is with the School of Physics and Information Technology, Shaanxi Normal University, Xi'an 710119, China (e-mail: wteng@snnu.edu.cn)}
\thanks{C. Liu is with the State Key Laboratory of Networking and Switching Technology, Beijing University of Posts and Telecommunications, Beijing 100876, China (e-mail: chenxi.liu@bupt.edu.cn)}
\thanks{W. Feng is with the Department of Electronic Engineering, State Key Laboratory of Space Network and Communications, Tsinghua University, Beijing 100084, China (e-mail: fengwei@tsinghua.edu.cn)}}

\maketitle
\begin{abstract}
Artificial intelligence-generated content (AIGC) service provisioning in wireless edge networks involves two phases: content generation on edge servers and content transmission to mobile devices. In this paper, we take image generation as a representative application and propose a batch denoising framework, followed by a joint optimization of content generation and transmission, with the objective of maximizing the average AIGC service quality under an end-to-end service delay constraint. Motivated by the empirical observations that (i) batch denoising effectively reduces per-step denoising delay by enhancing parallelism and (ii) early denoising steps have a greater impact on generation quality than later steps, we develop the STACKING algorithm to optimize batch denoising. The STACKING operates independently of any specific form of the content quality function and achieves lower computational complexity. Building on the batch solution, we further optimize bandwidth allocation across AIGC services. Simulation results demonstrate the superior performance of our algorithm in delivering high-quality, lower-latency AIGC services.
\end{abstract}


\section{Introduction}
Artificial intelligence-generated content (AIGC) services refer to applying generative artificial intelligence (GenAI) models to provide personalized and dynamic content generation services like image and video generation \cite{10445209}.
With the ubiquity of mobile devices, the demand for high-quality and low-latency AIGC service provisioning in wireless edge networks is rapidly increasing \cite{10742103}. 
For example, integrating GenAI into wireless edge networks enables novel vehicular services, such as augmented reality road simulation \cite{10528244}. 
The complete process of AIGC services involves using GenAI models deployed on edge servers to generate content according to user requirements, followed by transmitting the generated results to mobile devices.

During the content generation phase, GenAI
models are primarily based on diffusion processes for mainstream image and video generation, producing content through multiple denoising steps. Increasing the number of denoising steps generally improves the generated content quality but also prolongs the generation delay \cite{2021Denoising}. To balance AIGC service quality and responsiveness, establishing a quantitative relationship between denoising steps and service quality is of importance.  
Besides, diffusion-based GenAI models involve iterative denoising with significant computational delay, while delivering generated content from edge servers to mobile devices introduces non-negligible transmission delay. Hence, there is a need for joint optimization of denoising quality and latency, supported by efficient allocation of heterogeneous communication and computational resources \cite{10804565}.

To explore the relationship between denoising steps and AIGC service quality, some preliminary studies have been conducted \cite{10683477,liu2024two}. They either use simple utility functions that only approximate the relationship, or fit model-specific complex mathematical forms, which hinder effective optimization of service quality.
Besides, existing research has integrated batch processing with radio resource allocation to support AI inference in wireless edge networks \cite{9843917,10038543,10545324}. Batch processing reduces per-service computational delay by grouping multiple services into a single batch for concurrent processing, amortizing memory access time. However, 
unlike AI inference considered in previous works, diffusion-based AIGC involves multiple denoising steps, i.e., a sequence of dependent inference tasks, which makes the conventional batching approaches inapplicable. Thus, how to schedule denoising tasks across batches while considering radio resource allocation, in order to balance the quality and responsiveness of multiple AIGC services, remains an open problem.

In this paper, we consider a batch denoising framework for multi-user AIGC services in wireless edge networks, using image generation as a representative application. Aiming to maximize average generation quality subject to the end-to-end service delay, we formulate a joint generation-and-transmission optimization problem by optimizing batch denoising and bandwidth allocation. We accurately model the relationship between the number of denoising steps of denoising diffusion implicit model (DDIM) \cite{2021Denoising} and content quality, measured by the Fréchet Inception Distance (FID). Inspired by enhanced parallelism of batch denoising to reduce per-step denoising delay and the greater impact of initial denoising steps on generation quality compared to later steps, we design the STACKING algorithm for batch denoising, which follows a clu{\emph{st}}ering-p{\emph{ack}}ing-batch{\emph{ing}} process. The STACKING algorithm is agnostic to the specific properties of the content quality function and reduces computational complexity effectively.
Building on the batch denoising solution, we further optimize bandwidth allocation among AIGC services using a straightforward particle swarm optimization (PSO) algorithm.
Simulation results demonstrate the superiority of the proposed algorithm for high-quality and low-latency AIGC service provisioning.

The remainder of this paper is organized as follows. Section II presents the system model and problem formulation, followed by the algorithm design in Section III. Simulation results are showcased in Section IV. Finally, we draw conclusions in Section V.

\section{System Model and Problem Formulation}
We consider a single-cell network scenario in which an edge server provides AIGC services to \(K\) mobile devices, indexed by \(\mathcal{K} = \{1,  \ldots, K\}\). 
On the server, a pre-trained GenAI model, such as DDIM, is deployed and shared among $K$ devices to support their AIGC services within diverse end-to-end delay requirements. A complete AIGC service consists of two stages: 1) content generation on the server; 2) content transmission from the server to the device. Then, we elaborate on the system models for these two stages. Note that we will use the terms “device” and “service” interchangeably throughout the paper, owing to their one-to-one correspondence.

\subsection{Content Generation Model} 
We adopt batch denoising to enable fast content generation. For service \(k\), the number of denoising steps is denoted by \(T_k\), with the corresponding set defined by \(\mathcal{T}_k = \{1, \ldots, T_k \}\). Each denoising step is treated as an individual denoising task. During batch denoising, the server groups the denoising tasks of  \(K\) services into \(N=\sum_{k\in\mathcal{K}} T_k\) batches, indexed by the set \(\mathcal{N} = \{1,\ldots, N\}\). The $N$ batches are processed sequentially, with start times satisfying \(t_1 \leq t_2 \leq \cdots \leq t_N\), where \(t_n\) denotes the start time of the \(n\)-th batch.

Let \(x_{k,n}^{s}\) be a binary variable indicating whether the \(s\)-th denoising task of service \(k\) is assigned into the \(n\)-th batch: 
\begin{equation}
x_{k,n}^s \in \{0, 1\}, \forall k \in \mathcal{K}, \forall n \in \mathcal{N}, \forall s \in \mathcal{T}_k.
\label{eq1}
\end{equation}
If assigned, \(x_{k,n}^{s}=1\); otherwise, \(x_{k,n}^{s}=0\). To achieve the expected content quality, each denoising task of a service must be executed exactly once:
\begin{equation}
\sum_{n \in \mathcal{N}} x_{k,n}^s = 1, 
\forall k \in \mathcal{K}, \forall s \in \mathcal{T}_k.
\label{eq2}
\end{equation}
For the \(n\)-th batch, its size, i.e., the number of assigned denoising tasks, is calculated as
\begin{equation}
X_{n}=\sum_{k\in \mathcal{K}}\sum_{s\in \mathcal{T}_k}x_{k,n}^{s},
\label{eq3}
\end{equation}
which directly determines the denoising delay of \(n\)-th batch.

\begin{figure}[t]
    \centering
    \subfloat[]{
        \includegraphics[width=0.45\textwidth]{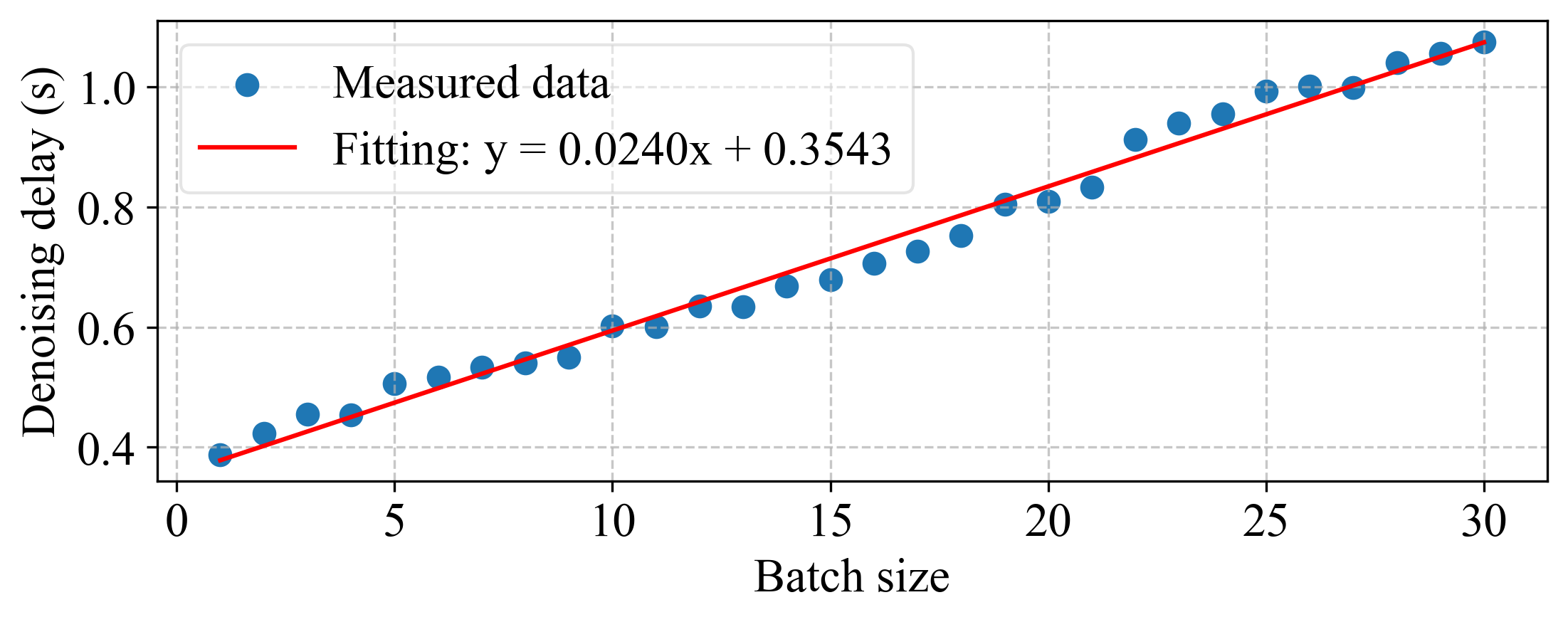}
        \label{fig1}
    }
    \vspace{0 cm} 
    \subfloat[]{
        \includegraphics[width=0.45\textwidth]{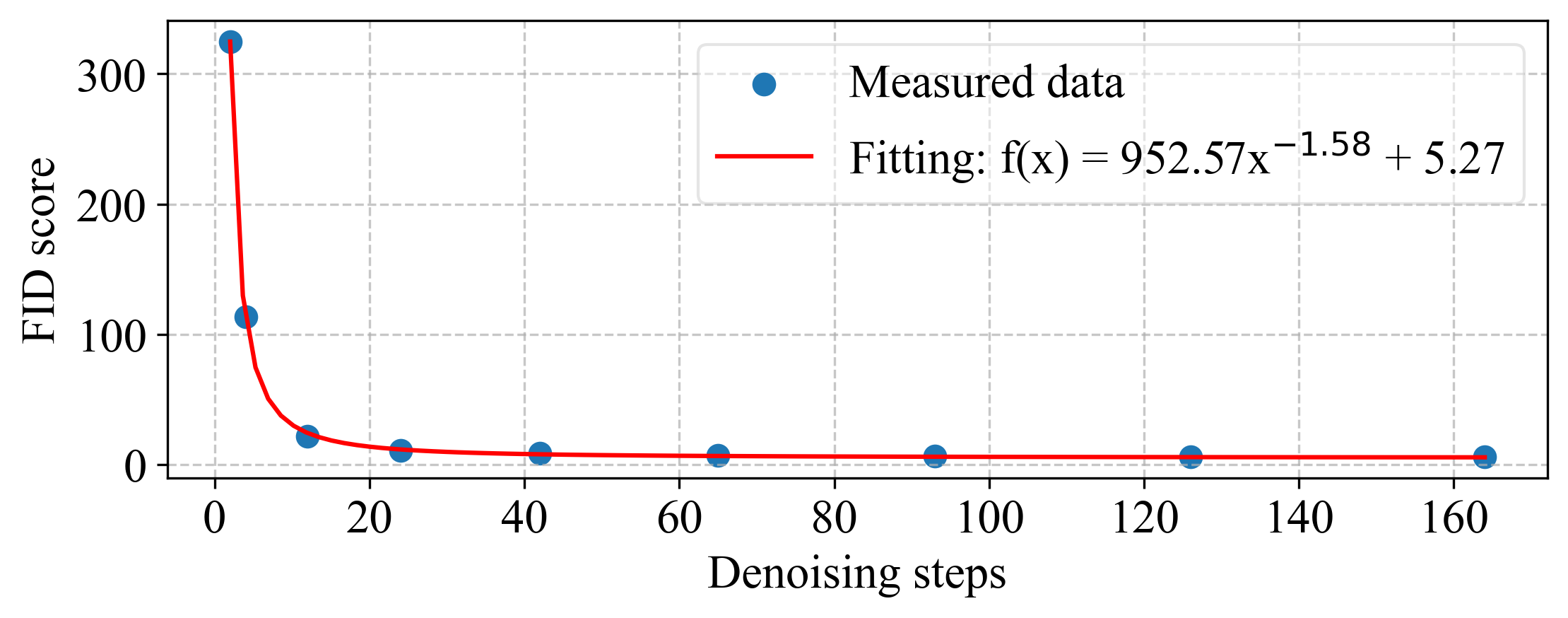}
        \label{fig2}
    }
    \caption{Measured data and fitting curves: (a) Denoising delay vs. batch size; (b) FID score vs. denoising steps.}
    \label{fig:fitting}
\end{figure}

It is observed from Fig.~\ref{fig1}, the {denoising delay} of the \(n\)-th batch can be modeled as
\begin{equation}
g(X_n)=aX_n+b\|X_n\|_0,
\label{eq4}
\end{equation}
where the parameters \(a\) and \(b\) are constants tailored to specific GenAI model and graphics processing unit (GPU) hardware, and \(\| \cdot \|_0\) is \(\ell_0\)-norm. In Fig.~\ref{fig1}, the measured values, obtained from the DDIM model pretrained on CIFAR-10 and run on an NVIDIA GeForce RTX 3050, are well fitted by a linear function with $a=0.0240$ and $b=0.3543$.

On this basis, the completion time of all denoising tasks for service \(k\), which also represents the content generation delay for service \(k\), can be expressed as
\begin{equation}
D_{k}^{\textnormal{cg}}=\sum_{n\in N}x_{k,n}^{T_{k}}(t_{n}+g(X_{n})).
\label{eq5}
\end{equation}
Due to the sequential execution of batches and the dependencies among denoising tasks within a single service, the following two constraints must be satisfied. First, the \((n + 1)\)-th batch cannot be  processed until the \(n\)-th batch is completed: 
\begin{equation}
t_{n} + g(X_{n}) \leq t_{n+1}, \forall n \in \mathcal{\tilde{N}},
\label{eq6}
\end{equation}
with \(\mathcal{\tilde{N}}=\{1, \ldots, N-1\}\). Second, for service \(k\), the \((s+1)\)-th denoising task cannot be processed until the \(s\)-th denoising task is completed:
\begin{equation}
\sum_{n \in \mathcal{N}} x_{k,n}^{s} \left(t_{n}+ g(X_{n})\right) \leq \sum_{n \in \mathcal{N}} x_{k,n}^{s+1} t_{n}, \forall k\in\mathcal{K}, \forall s \in \mathcal{\tilde{T}}_k,
\label{eq7}
\end{equation}
with \(\mathcal{\tilde{T}}_k=\{1, \ldots, T_{k}-1\}\).

\subsection{Content Transmission Model}
Once the content is generated, it is then transmitted back to the mobile device from the edge server. For simplicity, we consider frequency non-selective channels and unchanged channel conditions during the content transmission \cite{OFDM_ChaoXu}. 
Let \(B\) denote the total bandwidth and \(B_k\) denote the bandwidth allocated to device \(k\). Then, we can calculate the transmission rate from the edge server to device \(k\) as:
\begin{equation}
r_{k} = B_k \eta_k,
\label{eq8}
\end{equation}
where \(\eta_k=\log_2\left(1 + \frac{\bar{p} h_{k}}{N_{0}}\right)\) denotes the spectral efficiency. Specifically, \(\bar{p}\) denotes the transmit power per unit bandwidth of the edge server, \(h_{k}\) represents the channel gain between the server and device $k$, and \(N_{0}\) is the noise power spectral density. The sum of the bandwidth allocated to all devices cannot exceed the total bandwidth:
\begin{equation}
\sum_{k \in \mathcal{K}} B_k \leq B,
\label{eq9}
\end{equation}
and the bandwidth allocated to device $k$ satisfies:
\begin{equation}
    0 \leq B_k \leq B,\forall k\in\mathcal{K}.
    \label{eq10}
\end{equation}

Further, the content transmission delay from the server to device $k$ can be given by:
\begin{equation}
D_{k}^{\textnormal{ct}} = \frac{S}{r_{k}}.
\label{eq11}
\end{equation}
Here, \(S\) is the content size, which is identical across all services as the content is generated by the same GenAI model.

\subsection{Problem Formulation}
By summing the content generation delay in \eqref{eq5} and transmission delay in \eqref{eq11}, we can give the end-to-end delay for service \(k\) by:
\begin{equation}
D_{k}^{\textnormal{e2e}} = D_{k}^{\textnormal{cg}} + D_{k}^{\textnormal{ct}},
\label{eq12}
\end{equation}
which cannot exceed the predefined deadline $\tau_k$, as follows:
\begin{equation}
D_{k}^{\textnormal{e2e}} \leq \tau_k, \forall k \in \mathcal{K}.
\label{eq13}
\end{equation}

Under the deadline constraints, the AIGC service quality maximization problem can be formulated as
\begin{align}
\text{(P0)} \min_{T_k, t_n, x_{k,n}^s, B_k} &\quad \frac{1}{K}\sum_{k \in \mathcal{K}} Q_k \nonumber \\
\text{s.t.} \quad\quad & \quad \eqref{eq1},\eqref{eq2},\eqref{eq6},\eqref{eq7}, \nonumber \\ 
& \quad \eqref{eq9},\eqref{eq10},\eqref{eq13}, \nonumber
\end{align}
where the objective is to optimize the content generation process via batch denoising (including the denoising steps, batch start times, and task-batch assignments), and the content transmission process via bandwidth allocation. In addition, \eqref{eq1}, \eqref{eq2}, \eqref{eq6}, and \eqref{eq7} put restrictions on the batch denoising; \eqref{eq9} and \eqref{eq10} regulate the bandwidth allocation; and  \eqref{eq13} ensures the deadline constraints are satisfied. 

Problem (P0) is hard to solve due to the following reasons:
\begin{itemize}
    \item The content generation quality $Q_k$ for service $k$ is an implicit function of the number of denoising steps, generally exhibiting a complex form. In Fig. \eqref{fig2}, we use the FID score to evaluate the image generation quality and show its relationship with the number of denoising steps, based on image generated by DDIM pretrained on CIFAR-10. As the denoising steps increase, the FID score drops sharply at first and then gradually levels off, indicating improved image quality. A power-law function is employed to accurately fit the measured data.
    \item The content generation delay in \eqref{eq5} depends on the task-batch assignment, where the presence of \(\ell_0\)-norm and product terms introduces significant complexity. 
\end{itemize}
In the next section, we leverage the characteristics of the measured data in Figs.~\ref{fig1} and \ref{fig2}, to design an effective algorithm to solve problem (P0). 

\section{Algorithm Design}
\subsection{Problem Reformulation}
To make problem (P0) tractable, we recast it as 
\begin{align}
\text{(P1)} \min_{B_k} &\quad Q^*(B_1,\ldots,B_K)\nonumber \\
\text{s.t.}  &\quad \eqref{eq9},\eqref{eq10}, \nonumber  
\end{align}
where \(Q^*(B_1,\ldots,B_K)\) denotes the optimal value of the following problem \cite{C-RAN-JSAC}:
 \begin{align}
\text{(P2)}\min_{T_k, t_n, x_{k,n}^s} &\frac{1}{K}\sum_{k \in \mathcal{K}} Q_k \nonumber \\
\text{s.t.}\quad
&\eqref{eq1},\eqref{eq2},\eqref{eq6},\eqref{eq7}, \nonumber \\
&D_{k}^{\textnormal{cg}} \leq  \tau^{\prime}_k \triangleq \tau_k - D_{k}^{\textnormal{ct}}, \forall k \in \mathcal{K} \label{eq14}.
\end{align}
Note that, problem (P2) is an optimization problem focused on batch denoising, with bandwidth allocation fixed. Once problem (P2) is solved, its optimal value can be substituted into problem (P1), thereby reducing problem (P1) to a bandwidth allocation problem. Based on this decomposition, we solve the two problems successively.

\subsection{Batch Denoising Optimization}
Observed from Fig.~\ref{fig1}, parameter $b$ is larger than parameter $a$, indicating that the overhead of the GPU loading the model from memory is significantly higher than the computational delay caused by increasing the batch size. Therefore, the first idea of batch denoising optimization is: to maximize the number of denoising tasks in each batch as much as possible, i.e., \(X_n\), thereby improving computational efficiency. However, the increase in \(g(X_n)\) results in some strict delay-constrained services being unable to complete sufficient denoising ones, which limits the content generation quality. As shown in Fig.~\ref{fig2}, the initial denoising steps have a far greater impact on image quality than subsequent steps. Therefore, the second idea of batch denoising optimization is: to balance the number of denoising steps for different services as much as possible. 

Inspired by the above two ideas, we design a STACKING algorithm to solve (P2) with the aid of an auxiliary variable $T^{*}$, which is defined as the expected number of denoising tasks for each service. In detail, the STACKING algorithm consists of the following three steps. 

\subsubsection{Clustering}
At the beginning of the $n$-th batch, we record the current number of completed denoising tasks for service \(k\) as \(T_{k}^{\textnormal{c}}=T_{k}\), and update its remaining delay as  
\begin{equation}
\tau^{\prime}_{k} = \tau^{\prime}_{k} - g(X_{n-1}).
\end{equation}  
The current maximum number of denoising tasks that service \(k\) can still complete is computed as
\begin{equation}
T_{k}^{\textnormal{e}} = \left\lfloor \frac{\tau^{\prime}_{k}}{a + b} \right\rfloor.
\end{equation}  
Thus, the ideal maximum number of denoising tasks that each service \(k\) can finally complete is given by  
\begin{equation}
T_{k}^{\prime} = T_{k}^{\textnormal{c}} + T_{k}^{\textnormal{e}}.
\end{equation}  
By comparing \(T^*\) and \(T_k^\prime\), all services are partitioned into two clusters: one is
\begin{equation}
    \mathcal{F} = \{k|T_{k}^{\prime} \leq T^*,\forall k\in\mathcal{K}\}
\end{equation}
and the other is \(\mathcal{K} \setminus \mathcal{F}\). To facilitate the description, all services in $\mathcal{K}$ are sorted in ascending order of $T_k^{\prime}$.

\subsubsection{Packing}
After clustering, a packing procedure is performed to group services from the two clusters for subsequent batching. Since batching always prioritizes services with smaller \(T_{k}^{\prime}\), we design the packing strategy based on the following two cases.

Case 1:
When $\mathcal{F} \neq \emptyset$, we prioritize services in $\mathcal{F}$, which have strict delay requirements, and pack them together by setting $X_n=|\mathcal{F}|$. However, if \(|\mathcal{F}|\) is small, this setting postpones the processing of services in \(\mathcal{K} \setminus \mathcal{F}\), potentially reducing the number of denoising tasks that can be completed within their deadlines.
Thus, we appropriately increase $X_n$ to simultaneously pack some services with stricter delay requirements in \(\mathcal{K} \setminus \mathcal{F}\), and set $X_n$ as
\begin{equation}
X_n=\max\left\{|\mathcal{F}|,\min\left\{|\mathcal{K|}, \left\lfloor \frac{\tau^{\textnormal{min}}-bT^{\textnormal{e(max)}}}{aT^{\textnormal{e(max)}}} \right\rfloor\right\}\right\},
\label{xn1}
\end{equation}
in which the \(\left\lfloor \cdot \right\rfloor\) term ensures \(T_{k}^{\textnormal{e}}(aX_n+b) \leq \tau_k^{\prime}\) holds for all \(k \in \mathcal{F}\). This means that some services in $\mathcal{K} \setminus \mathcal{F}$ can be packed without reducing the number of completed denoising task for each service in \(\mathcal{F}\). In addition, \(T^{\textnormal{e(max)}}\) and \(\tau^{\textnormal{min}}\) denote the maximum value of \(T_{k}^{\textnormal{e}}\) and the minimum value of \(\tau_{k}^{\prime}\) respectively, within \(\mathcal{F}\).

Case 2:
When $\mathcal{F} = \emptyset$, no services require priority processing and \(X_n\) is expected to be as large as possible, with an upper bound of \(|\mathcal{K}|\).
However, an excessively large \(X_n\) may significantly reduce the number of denoising tasks of services with strict delay requirements in \(\mathcal{K} \setminus \mathcal{F}\). Hence, we set \(X_n\) as 
\begin{equation}
X_n=\min\left\{|\mathcal{K}|,\left\lfloor \frac{(a+b)T^{\prime\textnormal{(min)}}-bT^*}{aT^*} \right\rfloor\right\} ,
\label{xn2}
\end{equation}
to maximize the batch size while ensuring \(T_{k}^{\prime} \geq T^*\) for all \( k \in \mathcal{K} \setminus \mathcal{F}\). This condition is enforced by the \(\left\lfloor \cdot \right\rfloor\) term in \eqref{xn2}, where \(T^{\prime\textnormal{(min)}}\) denotes the minimum value of \(T_{k}^{\prime}\) in \(\mathcal{K} \setminus \mathcal{F}\), making \((aX_n+b)T^{*} \leq (a+b)T_{k}^{\prime}\) hold for all \( k \in \mathcal{K} \setminus \mathcal{F}\).

\begin{algorithm}[t]
\caption{STACKING for solving problem (P2) \label{alg:bp}}
\begin{algorithmic}[1]
\STATE Initialize \(Q=\infty\);
\FOR{$T^*=1,...,T^{*\textnormal{max}}$}
\STATE Initialize \(n=1\), $t_n=0$, $s=0$, and \(T_k=0\) for any service $k$;
\WHILE{\(\mathcal{K}\neq \emptyset\)}
    \STATE Execute the clustering-packing-batching process, update $s\gets s+1$, $x_{k,n}^s=1$, and $T_k\gets s$ for any service $k$ in the $n$-th batch, and remove services violating their delay requirements from $\mathcal{K}$; 
    \STATE \(n\gets n+1\) and $t_n \gets t_n +g(X_n)$;
\ENDWHILE
\STATE Based on $T_k$, calculate the objective function value of problem (P2) as \(Q'\);
\IF{\(Q'< Q\)}
    \STATE \(Q\gets Q'\), \(T_k^* \gets T_k\), {\color{black}\(t_n^*\gets t_n\)}, and \(x_{k,n}^{s*}\gets x_{k,n}^s\);
\ENDIF
\ENDFOR
\RETURN $Q$, \(T_k^*\), \(t_n^*\), and \(x_{k,n}^{s*}\).
\end{algorithmic}
\end{algorithm}

\begin{figure*}[t] 
    \centering     
    \subfloat[]{%
    \includegraphics[width=0.32\textwidth]{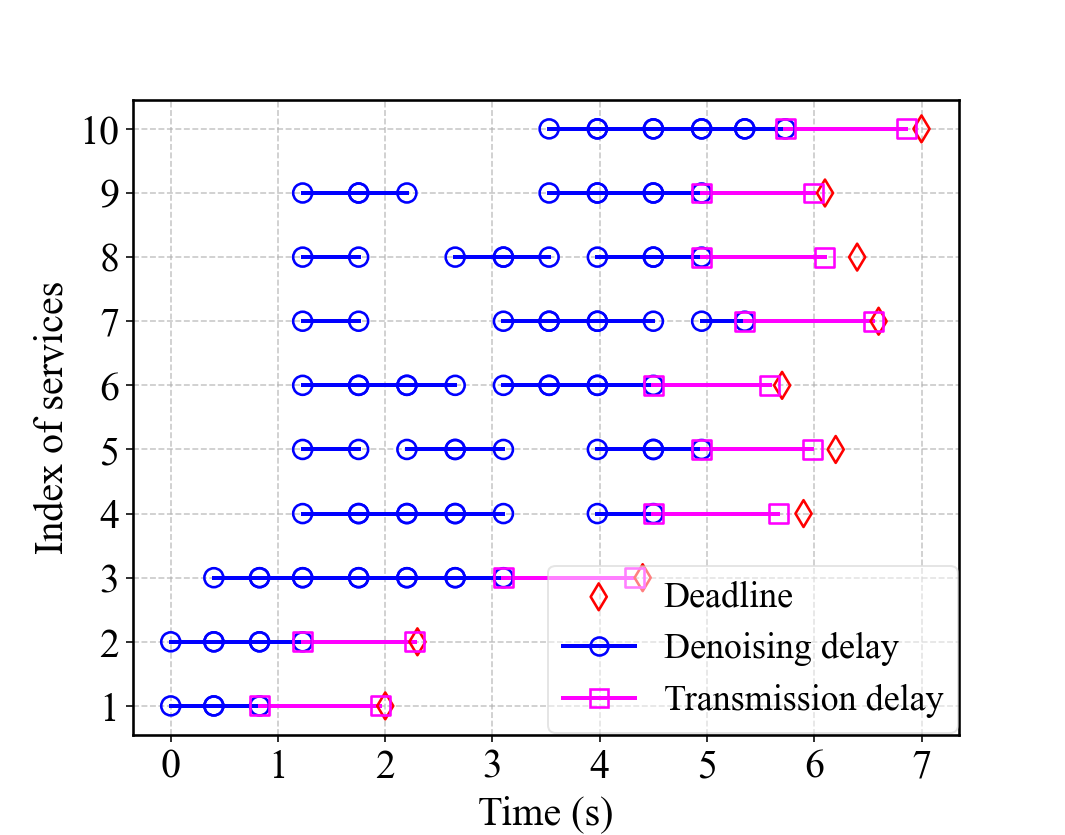} 
        \label{fig3}%
    } 
    \subfloat[]{%
        \includegraphics[width=0.32\textwidth]{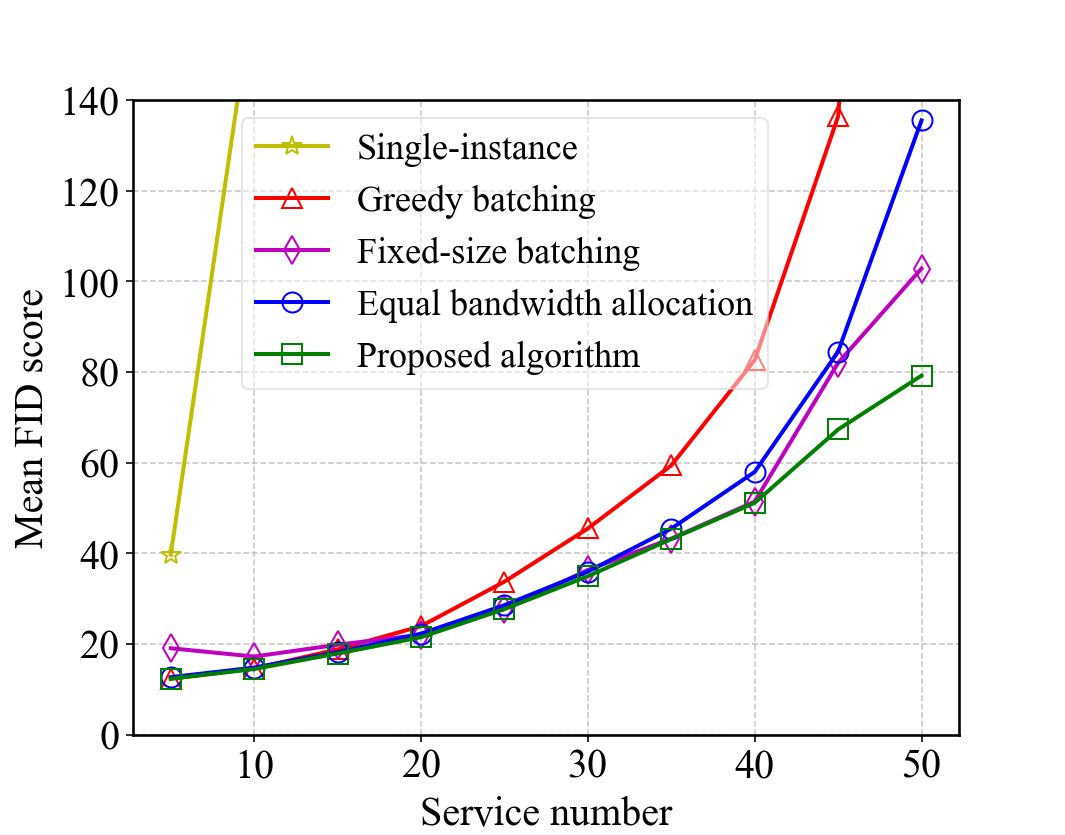}
        \label{fig4}%
    } 
    \subfloat[]{%
        \includegraphics[width=0.32\textwidth]{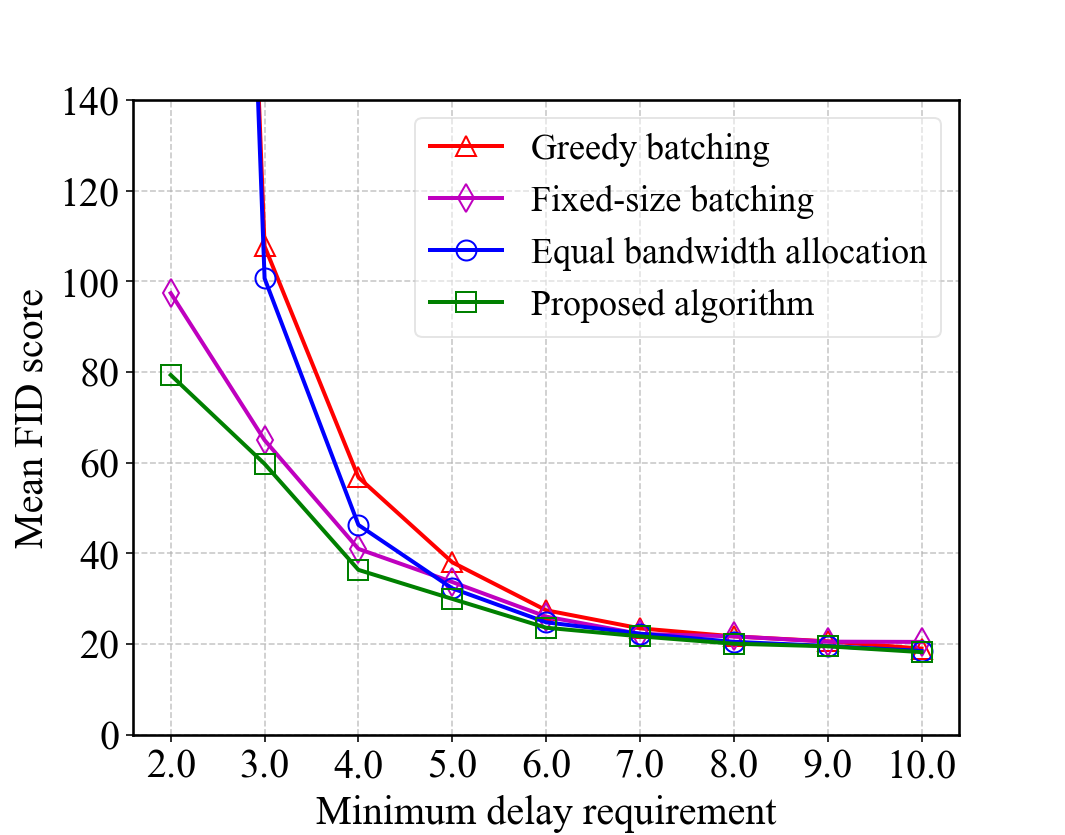}
        \label{fig5} 
    }
    \caption{Simulation results: (a) End-to-end delay illustration; (b) Mean FID sore vs. service number; (c) Mean FID score vs. minimum delay requirement.} 
    \vspace{-0.4cm}
    \label{fig:performance}                               
\end{figure*} 

\subsubsection{Batching}
The latest denoising tasks of the first $X_n$ services in the set $\mathcal{K}$ are packed into the $n$-th batch.
If the remaining delay \(\tau^{\prime}_k\) of any packed service \(k\) is less than \(g(X_n)\), the service is considered to have completed all of its denoising tasks. Consequently, its denoising task is removed from the batch and \(X_n\) is decreased by 1. The remaining tasks are batched for the denoising, along with updated \(g(X_n)\) and \(t_n\). Meanwhile, for each service $k$ in the $n$-th batch, its denoising step index \(s\) is increased by 1, \(x_{k,n}^{s}=1\) is set, and $T_k$ is updated to $s$. After the $n$-th batch denoising, services that violate their delay requirements are removed from $\mathcal{K}$ to prevent them from being processed in subsequent batches. 

The above clustering-packing-batching process is repeated until \(\mathcal{K}\) becomes empty, indicating that the denoising tasks of all AIGC services have been completed. Furthermore, the optimal value of $T^*$ depends on the distribution of delay requirements across all services. To determine this value, we perform a search within a predefined range and select the value that minimizes the objective function of (P2). The complete procedure is summarized in Algorithm \ref{alg:bp}.

\subsection{Bandwidth Allocation Optimization}
After solving problem (P2), we turn our focus on problem (P1). Specifically, we apply the PSO algorithm \cite{488968} to iteratively search for the optimal \(B_k\) that maximizes the objective function of problem (P1). 

\section{Simulation Results}
Unless otherwise specified, we consider \(K = 20\) devices, each with an AIGC service request and a delay requirement uniformly distributed between 7 and 20 seconds. The DDIM model, pretrained on CIFAR-10, is employed to generate the required content (i.e., images). For content transmission, the total available bandwidth \(B\) is set as 40 KHz and the spectral efficiency is assumed to range between 5 and 10 bit/s/Hz.
The following baselines are considered for comparisons:
\begin{itemize}
\item Single-instance scheme \cite{Asy_MEC}: All services are sorted in ascending order of delay requirements. Then, the server processes their denoising tasks sequentially, without batching, and removes a service once its delay requirement is violated.
\item Greedy batching scheme: The server groups denoising tasks from all services into a batch and processes them in parallel. Once a service exceeds its delay constraint, the server terminates its denoising process. 
\item Fixed-size batching scheme: The server uses a fixed batch size, set to \(\lfloor K/2 \rfloor\). It prioritizes services with tighter delay constraints and discards those that violate their deadlines. When the number of remaining services is smaller than the batch size, the server reduces the batch size to match.
\item Equal bandwidth allocation scheme: Equal bandwidth is assigned to each device, and Algorithm \ref{alg:bp} is used for batch denoising.
\end{itemize}
Note that, the bandwidth allocation strategy used in our algorithm is also applied to the single-instance, greedy batching, and fixed-size batching schemes. 

Fig.~\ref{fig3} shows the end-to-end delay of 10 AIGC services under the proposed algorithm. 
We can observe that services with tighter delay requirements are prioritized for processing. 
When only services with looser delay constraints remain, the server processes as many of them as possible to maximize the computational resource utilization. Since the initial denoising steps contribute most significantly to content quality, services with similar delay requirements tend to be assigned a similar number of denoising steps to optimize overall service quality. Moreover, the bandwidth is reasonably allocated so that most services complete their content transmission close to the deadline, allowing more time to be reserved for content generation and thus improving overall service quality.

From Fig.~\ref{fig4}, we observe that an increase in the number of services results in a higher mean FID score, indicating degraded average AIGC service quality. The single-instance scheme struggles to support multi-user AIGC services, highlighting the necessity of batch denoising. Both the greedy batching and fixed-size batching schemes fail to effectively allocate computational resources under increasing service loads, leading to a rapid rise in mean FID score and even causing service outages. 
Furthermore, the benefits of bandwidth allocation become more pronounced as the number of services grows. This highlights the importance of rational bandwidth allocation, especially under constrained bandwidth conditions.

Fig.~\ref{fig5} illustrates the mean FID score versus the minimum delay requirement, where the maximum delay requirement is fixed at 20 seconds. The proposed algorithm consistently achieves the lowest mean FID score. Notably, the smaller the minimum delay requirement, the greater the performance gain. 
Compared with the greedy batching and fixed-size batching schemes, the proposed algorithm performs better under strict delay constraints due to its flexible batch denoising strategy. Furthermore, in comparison with the equal bandwidth allocation scheme, the proposed algorithm provides higher-quality AIGC service particularly when the minimum delay requirement becomes tight. 

\section{Conclusions}
In this paper, we have proposed a batch denoising framework to support multi-user AIGC services with heterogeneous delay requirements. Motivated by the insights from Figs. \ref{fig1} and \ref{fig2}, we have proposed the STACKING algorithm to optimize batch denoising without relying on a specifically defined content quality function. Building on the batch denoising results, we have further optimized bandwidth allocation to better balance AIGC service quality and responsiveness. Simulation results have been presented to demonstrate the superiority of the proposed algorithm.

\normalem
\bibliographystyle{IEEEtran}
\bibliography{AIGC}

\end{document}